\documentclass[twocolumn,amsmath,aps,fleqn]{revtex4}

\usepackage{amssymb}
\usepackage{amsmath}
\usepackage{epsfig}
\usepackage{subfigure}
\usepackage{mathrsfs}

\newcommand{\be}{\begin{equation}}
\newcommand{\ee}{\end{equation}}
\newcommand{\bea}{\begin{eqnarray}}
\newcommand{\eea}{\end{eqnarray}}

\begin{document}

\title{Eddington's theory of gravity and its progeny.}
\author{M\'aximo Ba\~nados\footnote{JS Guggenheim Memorial Foundation Fellow}}
\affiliation{P.Universidad Cat\'olica de Chile, Av. Vicuna Mackenna 4860,
Santiago, Chile\\
Astrophysics, University of Oxford, DWB, Keble Road, Oxford, OX1 3RH, UK}
\author{Pedro G. Ferreira}
\affiliation{Astrophysics, University of Oxford, DWB, Keble Road, Oxford,
OX1 3RH, UK}


\begin{abstract}
We resurrect Eddington's proposal for the gravitational action in the
presence of a cosmological constant and extend it to include matter fields.
We show that the Newton-Poisson equation is modified in the presence of sources and that charged black holes show great similarities with those arising in Born-Infeld electrodynamics coupled to gravity.  When we consider homogeneous and isotropic space-times we find that there is
a minimum length (and maximum density) at early times, clearly
pointing to an alternative theory of the Big Bang. We thus argue that the
modern formulation of
Eddington's theory, Born-Infeld gravity, presents us with a novel,
non-singular description of the Universe.
\end{abstract}

\maketitle

\section{Introduction}
The Einstein-Hilbert action has been the mainstay of gravitational theory
for almost a century. It
can be expressed as $
S_{EH}\equiv\frac{1}{2}\int
d^4x\sqrt{|g|}(g^{\alpha\beta}R_{\alpha\beta}-2\Lambda)$
where $g_{\alpha\beta}$ is the metric of space-time (and $|g|$ is its
determinant), $R_{\alpha\beta}$ is
the Ricci tensor of that metric,
$\Lambda$  is the cosmological constant (in this letter we will work in
Planck units $8\pi G=1$). An intriguing, alternative proposal for the gravitational action was
proposed by Eddington in 1924 \cite{Eddington}. He suggested that, at least
in free, de-Sitter space, the fundamental field should be  the
connection, $\Gamma^{\alpha}_{\beta\gamma}$ and the relevant action should
be
\begin{eqnarray}
S_{Edd}=2\kappa\int d^4x\sqrt{|R|} \label{sedd}
\end{eqnarray}
where $\kappa$ is a constant with inverse dimensions to that of $\Lambda$ and $|R|$ is the
determinant of $R_{\mu\nu}$.  Note that $R_{\alpha\beta}$ is constructed {\it solely} from the connection. (Here and in the rest of this Letter, $R_{\mu\nu}(\Gamma)$ represents the symmetric part of the Ricci tensor build with the connection.)

Varying $S_{Edd}$, integrating by parts and eliminating a vanishing trace we obtain
\begin{eqnarray}
\nabla_\alpha(2\kappa \sqrt{|R|}R^{\mu\nu})=0 \nonumber
\end{eqnarray}
where $\nabla$ is the covariant derivative defined in terms of
$\Gamma^{\alpha}_{\beta\gamma}$ and $R^{\mu\nu}$ is the inverse of the Ricci tensor.
This equation can be partially solved if we define a new rank-2 tensor
$q_{\mu\nu}$ such that  $\nabla_\alpha (\sqrt{|q|} q^{\mu\nu})=0$.  The field equations for this
theory then become
\begin{eqnarray}
2\kappa\sqrt{|R|}R^{\mu\nu}=\sqrt{|q|}q^{\mu\nu}
\end{eqnarray}
which can be rewritten as the Einstein field equations if we equate
$q_{\alpha\beta}$ with $g_{\alpha\beta}$ and $\kappa$ with $\Lambda^{-1}$.
Hence, Eddington's action should be a viable,
alternative starting point to General Relativity. In fact they can be viewed
as being dual to each other- while $S_{EH}$ is proportional to $\Lambda$,
$S_{Edd}$ is inversely proportional to $\Lambda$ and one can imagine that
they should be useful in different regimes.

Eddington's theory of gravity is incomplete in that it doesn't include matter. There have been subsequent attempts at coupling matter to $\Gamma$ in the original Eddington spirit \cite{kijowski}. The idea is as follows.  Start with a Palatini gravitational action coupled to matter $I[g,\Gamma,\Psi]$. Here $\Psi$ denotes all matter fields. The metric enters with no derivatives and it is, in principle \cite{comment}, possible to express $g_{\mu\nu}$ in terms of $\Gamma$ and $\Psi$ using its own equation of motion. Since this step is algebraic, one can replace this formula back into the action and obtain an ``affine variational principle" $I'[\Gamma,\Psi]$ depending only on the connection and matter fields. The action $I'[\Gamma,\Psi]$ derived by this procedure may be quite complicated  \cite{kijowski}, but it must be stressed that its dynamics is fully equivalent to the original metric theory. As a matter of fact, this procedure is most simple without matter and precisely maps the Palatini version of $S_{EH}$ into $S_{Edd}$.

In this Letter, we shall reconsider the problem of coupling Eddington action to matter without insisting neither on a purely affine action nor on a theory equivalent to Einstein gravity. We shall allow the metric to be present and switch to a Born-Infeld  \cite{BornInfeld} like structure, $\sqrt{|g_{\mu\nu}+\kappa R_{\mu\nu}|}$. Observe that for large $\kappa R_{\mu\nu}$ one reobtains the Eddington functional. A purely metric proposal can be found in \cite{DeserGibbons}. We shall focus on a Palatini formulation (with $g_{\mu\nu}$ and $\Gamma^{\mu}_{\ \alpha\beta}$ independent) of the gravitational action first proposed in \cite{Vollick}
\begin{eqnarray}
S_{BI}[g,\Gamma,\Psi]&=&\frac{2}{\kappa}\int d^4x \left[\sqrt{|g_{\mu\nu}+\kappa R_{\mu\nu}(\Gamma)|}-\lambda \sqrt{g}\right] + \nonumber\\
&& \ \ \ \ \ \  S_M[g,\Gamma,\Psi] \label{ebi}
\end{eqnarray}
where $\lambda$ is dimensionless. Note that $\lambda$ must be different from zero. For $\lambda=0$ the metric variation yields (with no matter) $\sqrt{|g+\kappa R|}\left[g+\kappa R\right]^{-1 \mu\nu}=0$
which clearly makes no sense. In \cite{Vollick} the matter fields were introduced in a non-conventional way inside the square root. Here we add matter in the usual way.

This action has all the correct limits and no pathologies.  For small values of $\kappa R $, the action (\ref{ebi}) reproduces the Einstein-Hilbert action with $\Lambda = (\lambda-1)/\kappa$. On the other hand for large values of $\kappa R$, the action approximates Eddington's.  Unlike the matter actions built in \cite{kijowski} and reviewed above, (\ref{ebi}) is {\it not} equivalent to the Einstein-Hilbert action and we shall observe interesting deviations especially for large curvatures when Eddington action dominates.

In vacuum with $S_M=0$, on the other hand, the action (\ref{ebi}) is equivalent to $S_{EH}$. This can be proven in two steps. With $S_M=0$ the equation of motion for $g_{\mu\nu}$ implies $g_{\mu\nu}= {\kappa  \over \lambda-1} R_{\mu\nu}$. We then replace the metric back into the action to obtain the Eddington action ${\kappa\lambda \over \lambda-1} \int \sqrt{|R|}$ which we already remarked is equivalent to $S_{EH}$. The case $\lambda=1$ is not degenerate although this particular proof fails.  For $\lambda=1$ the equivalence can easily be proven at the level of the equations of motion. This is to be contrasted with $f(R)$ theories \cite{Sotiriou}, either metric or Palatini, where deviations are observed even in vacuum.  We shall now consider (\ref{ebi}) with matter in different situations to explore it properties and we shall restrict ourselves to the case where matter {\it only} couples to the metric (couplings to $\Gamma$ may arise due to quantum gravitational corrections). See \cite{other} for other discussions.

The equations of motion for this theory are the following. Varying with respect to $g_{\mu\nu}$ one obtains,
\begin{eqnarray}
 \frac{\sqrt{|g+\kappa R|}}{\sqrt{|g|}}[(g+\kappa R)^{-1}]^{ \mu
\nu}&-& \lambda g^{\mu\nu} = -\kappa T^{\mu\nu} \label{field3}
\end{eqnarray}
Here $T^{\mu\nu}$ is the standard energy momentum tensor with indices raised with the metric $g_{\mu\nu}$.

The variation with respect to $\Gamma$ can be simplified by introducing an auxiliary metric $q_{\mu\nu}$ compatible with $\Gamma$. The equation of motion becomes
\begin{eqnarray}
q_{\mu\nu}=g_{\mu\nu}+\kappa R_{\mu\nu} \label{field4}
\end{eqnarray}
and $\Gamma^{\mu}_{\ \alpha\beta} = {1 \over 2} q^{\mu\sigma}(q_{\sigma\alpha,\beta} + q_{\sigma\beta,\alpha}- q_{\alpha\beta,\sigma}). $ Combining (\ref{field4}) and (\ref{field3}) one finds the equation
\begin{equation}\label{field5}
\sqrt{q}q^{\mu\nu} = \lambda \sqrt{g} g^{\mu\nu} - \kappa \sqrt{g}\, T^{\mu\nu}.
\end{equation}
Again, here we have that $q^{\mu\nu}$ is the inverse of $q_{\mu\nu}. $Equations (\ref{field4}) and (\ref{field5}) form a complete set and provide the simplest set of equations to study this theory.  The conservation equations for the matter fields are the same as in GR,  $T^{\mu\nu}_{\ \ ;\mu}=0$, where the covariant derivative here refers to the metric $g_{\mu\nu}$.  Although this conservation equation is not at all obvious from the equations of motion (\ref{field3}), it is nevertheless true because matter is covariantly coupled to $g_{\mu\nu}$.

 We can now rewrite equation
 (\ref{field3}) as
 \begin{eqnarray}
\sqrt{|g+\kappa R|}{\bf I}=\sqrt{|g|}[(1+\kappa\Lambda){\bf g}^{-1}
-\kappa {\bf g}^{-1}{\bf T}{\bf g}^{-1}][{\bf g}+\kappa {\bf R}] \nonumber
\end{eqnarray}
where, for the moment, we are writing the equation in the matrix format, ${\bf I}$ is $4\times 4$ identity
matrix. If we take the determinant of both sides and replace it in the field equation, we find
\begin{eqnarray}
{\bf g}+\kappa {\bf R}&=&|g|^\frac{1}{2}|(1+\kappa\Lambda){\bf g}^{-1}-\kappa {\bf g}^{-1}{\bf T}{\bf g}^{-1}|^\frac{1}{2}
\nonumber \\ & & \ \ \ \ \  \times [(1+\kappa\Lambda){\bf g}^{-1}-\kappa {\bf g}^{-1}{\bf T}{\bf g}^{-1}]^{-1} \nonumber
\end{eqnarray}
i.e. we can solve for $R_{\mu\nu}$ in terms of $g_{\mu\nu}$ and $T_{\mu\nu}$. Note that this
equation is generic. Recall however that $R_{\mu\nu}$ is a function of $\Gamma^{\alpha}_{\mu\nu}$ (and therefore $q_{\mu\nu}$).

Expanding the field equations to 2$^{\rm nd}$ order in $\kappa$ to find the 1$^{\rm st}$
order corrections to Einstein's equations
\begin{eqnarray}
R_{\mu\nu}\simeq\Lambda g_{\mu\nu}+T_{\mu\nu}-\frac{1}{2}Tg_{\mu\nu} +  \kappa \left[S_{\mu\nu}-\frac{1}{4}Sg_{\mu\nu}\right] \nonumber
\end{eqnarray}
where
$S_{\mu\nu}=T_\mu^{\ \alpha} T_{\alpha\nu} \nonumber-\frac{1}{2}TT_{\mu\nu}$
(note that $\Lambda$ does not contribute to $S$).  Combining this equation
with the second field equation, (\ref{field4}), we have the lowest order correction to Einstein gravity.

With the lowest order correction in hand, we can study the non-relativistic limit.  As expected, the Poisson equation is modified in this theory. Consider a time independent metric
$ds^2 = -(1+2\Phi)dt^2 + (1-2\Psi) d\vec{x}\cdot d\vec{x} \nonumber$
where $\Phi$ and $\Phi$ only depend on $\vec{x}$, and an energy momentum tensor $T^{\mu\nu}=\rho u^\mu u^\nu$. We linearize equations (\ref{field3}) and (\ref{field4}) keeping terms linear in $\Phi,\Psi$ and $\rho$.  The full set of linearized equations are solved by $\Phi=\Psi$ plus the modified Poisson equation,
\begin{equation}\label{}
\nabla^2 \Phi = -{1 \over 2}\rho - {1 \over 4}\kappa \nabla^2 \rho. \label{Poisson}
\end{equation}
We come up against a key characteristic of this theory- it reproduces
Einstein gravity precisely within the vacuum but deviates from it in the presence of sources. We expect it therefore to play a role in regions of high density, such as within a black hole or in the very early universe.

An obvious next step is to explore the inner structure of black holes. Let us consider spherically symmetric configurations of the form
\begin{equation}\label{es}
ds^2 = -\psi(r)^2 f(r) dt^2  + {dr^2 \over f(r)} + r^2d\Omega^2. 
\end{equation}
As we have just remarked, in vacuum, the action (\ref{ebi}) is fully equivalent to $S_{EH}$ with $\Lambda = (\lambda-1)/\kappa$. The Schwarzschild-dS metric with $\psi=1$ and $f=1-2m/r - \Lambda r^2/3$ is thus a solution to (\ref{ebi}) with no sources. New interesting effects show up when matter is present.

The simplest form of ``matter"  is an electromagnetic field (matter in the sense that $T^{\mu\nu}\neq 0$).  Let us add $-{1 \over  4 }\sqrt{g} F_{\mu\nu}F^{\mu\nu} $ to the action (\ref{ebi}).  The exact solution can be found and involves elliptic integrals. For simplicity we set $\lambda=1$ yielding an asymptotically flat geometry. The solution is:
\begin{eqnarray}
f(r) &=& \left( \int \!{\frac { \left(r^2- {q}^{2} \right) 
 \left( {r}^{4} - \kappa\,{q}^{2} \right) }{{r}^{4}\sqrt {{r}^{4}+\kappa\,{q}^{
2}}}}{dr}-2M \right){ \sqrt {{r}^{4}+\kappa\,{q}^{2}}\, r \over 
  {r}^{4}- \kappa\,{q}^{2}}
 \nonumber\\
\psi(r) &=& {\frac {{r}^{2}}{\sqrt {2\,{r}^{4}+2\,\kappa\,{q}^{2}}}}\nonumber\\
E(r) &=& {\frac {q}{\sqrt {{r}^{4}+\kappa\,{q}^{2}}}}\nonumber
\end{eqnarray} 
where $E(r)$ is the electric field. $M$ is the mass and $q$ the electric charge.  Observe the similitude with pure Born-Infeld electrodynamics. For $\kappa>0$ the electric field is everywhere regular. The metric functions and still singular at $r=0$ and at $r^2 = \sqrt{\kappa} q$. This singularity is however inside the horizon.

Strictly speaking, this solution represents the exterior solution to a charged object. In this sense, the $r=0$ singularity does not yet imply a singular solution. In order to explore the singularity structure of the action (\ref{ebi}) one needs to add normal matter and study a collapsing object. A simple model is an interior cosmology glued together with an exterior Schwarzschild metric. We shall discuss the details of this problem elsewhere but we can anticipate interesting conclusions: as we now show
the cosmology associated to (\ref{ebi}) predicts a maximum density and a singularity free Universe.

It has become clear that Eddington gravity can play a role in regions of high curvature
or density, the conditions one might expect in the early Universe. We therefore focus now on cosmology. We shall assume a homogeneous and isotropic metric,
\begin{equation}\label{}
ds^2 = -dt^2 + a(t)^2 d\vec{x}\cdot d\vec{x} \nonumber
\end{equation}
 coupled to an ideal fluid $T^{\mu\nu} = (p+\rho) u^\mu u^\nu + pg^{\mu\nu}$. As we remarked before, the fluid satisfies its standard conservation equation
$\dot \rho = -3 H ( P+\rho)$

\begin{figure}[htbp]
\begin{flushleft}
\vspace{-15pt}
\epsfig{figure=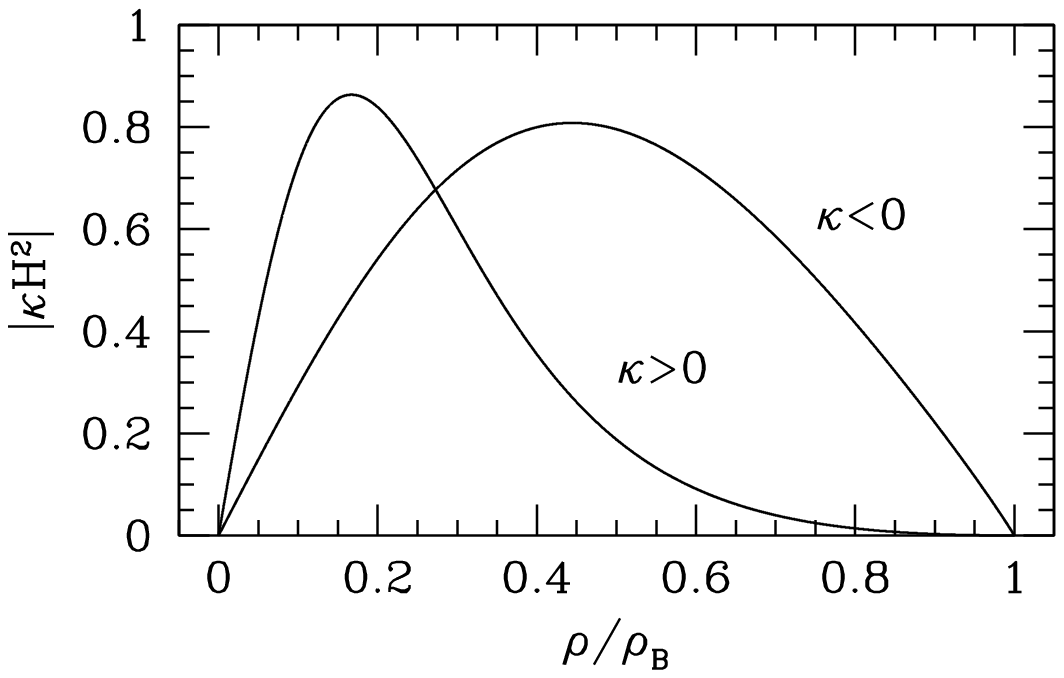,width=9cm}
\end{flushleft}
\vspace{-110pt}
\caption{The Hubble rate $H^2$ as a function of energy density, $\rho$, for a radiation filled
Universe with no cosmological constant and $|\kappa|=1$ We normalize the density by $\rho_B$ (for which
$H^2(\rho_B)=0$); for $\kappa>0$ we have $\rho_B=3/\kappa$ while for $\kappa<0$ we have
$\rho_B=-1/\kappa$. }
\label{H2rho}
\vspace{-10pt}
\end{figure}
\begin{figure}[htbp]
\begin{flushleft}
\vspace{-15pt}
\epsfig{figure=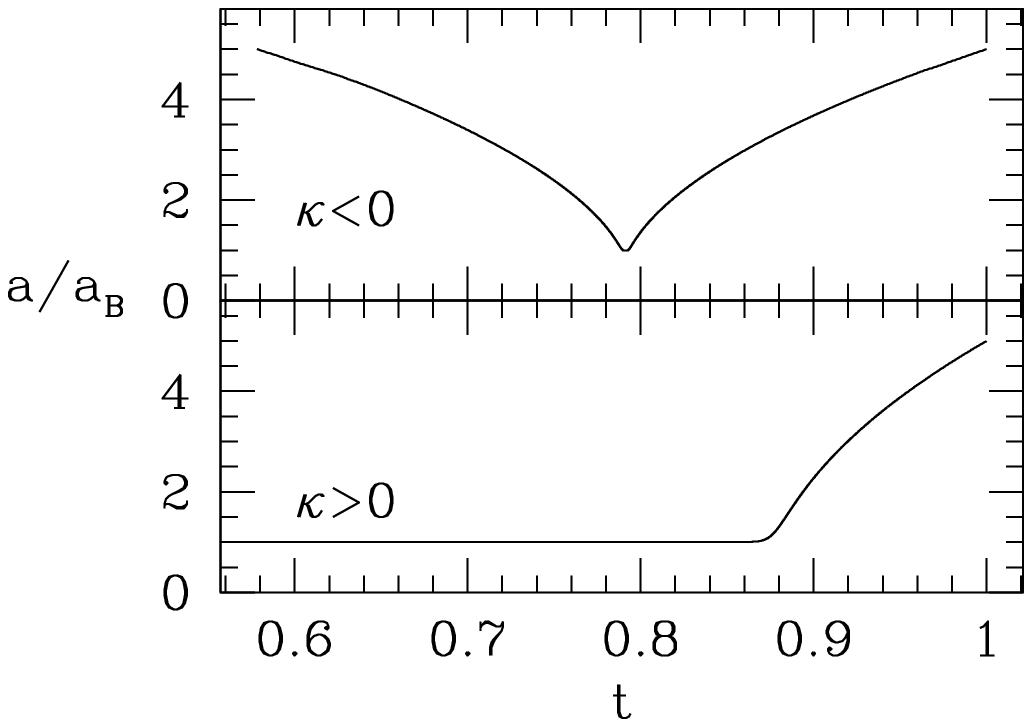,width=9cm}
\end{flushleft}
\vspace{-100pt}
\caption{The scale factor, a, normalized by the minimum length, a$_{\rm B}$, as a function of time, $t$ (in arbitrary units). In the top panel we show the existence of a bounce for $\kappa<0$ while in the bottom panel, the scale factor approaches the minimum length as $a\rightarrow -\infty$.}
\label{aoft}
\vspace{-10pt}
\end{figure}

We can assume that $q_{00}=-U$ and $q_{ij}=a^2V\delta_{ij}$ and use equations (\ref{field3}) and (\ref{field4}) to find
\begin{eqnarray}
U = \frac{D}{1+\kappa\rho_T} \ \  \mbox{\rm and} \ \
V = \frac{D}{1-\kappa P_T} \nonumber
\end{eqnarray}
where $D=\sqrt{(1+\kappa\rho_T)(1-\kappa P_T)^3}$, $\rho_T=\rho+\Lambda$ and
$P_T=P-\Lambda$. If we assume that $P=w \rho$, we can define
\begin{eqnarray}
F(\rho,\Lambda)&=&1-\frac{3(\kappa\rho_T+\kappa P_T)(1-w-\kappa\rho_T-\kappa P_T)}{4(1+\kappa \rho_T)(1-\kappa P_T)} \nonumber \\
G(\rho,\Lambda)&=&\frac{1}{\kappa}\left[1+2U-3\frac{U}{V}\right] \nonumber
\end{eqnarray}
to find the Friedman equations:
\begin{eqnarray}
H^2=\frac{1}{6}\frac{G}{F^2} \label{FRW}
\end{eqnarray}

If we assume $w=0$ (i.e. a dust filled universe with a cosmological constant), we can expand equation (\ref{FRW}) in terms of $\kappa \Lambda$ to find corrections to the
late time Friedman equations
\begin{eqnarray}
3H^2\simeq (\rho+\Lambda)+ \left[\frac{\rho^2}{\Lambda}-(\rho+\Lambda)\right]\kappa\Lambda+ {\cal O}[(\kappa\Lambda)^2] \nonumber
\end{eqnarray}
As expected, we find that we recover conventional Friedman cosmology at late times.

Let us now focus on the evolution of the scale factor at early times. Assuming radiation
domination, we have $\rho_T=\rho$, $P_T=P=\frac{1}{3}\rho$ and, definining
${\bar \rho}=\kappa\rho$ we find that equation (\ref{FRW}) becomes
\begin{eqnarray}
3H^2({\bar \rho})=&&\frac{1}{\kappa}\left[{\bar \rho}-1+\frac{1}{3\sqrt{3}}\sqrt{(1+{\bar \rho})(3-{\bar \rho})^3}\right]  \nonumber  \\ & & \ \ \ \ \ \ \times  \frac{(1+{\bar \rho})(3-{\bar \rho})^2}{(3+{\bar \rho}^2)^2} \nonumber
\end{eqnarray}
For small ${\bar \rho}$ we recover the conventional Friedman universe, $H^2\simeq \rho/3$ but
at high densities we come up against a novel effect: we find a stationary point, $H^2=0$ at  ${\bar \rho}=3$ (for $\kappa>0$)  and at  ${\bar \rho}=1$ (for $\kappa<0$).
The new stationary points correspond to a maximum density, $\rho_B$; in Figure (\ref{H2rho})  we plot the $H^2$ as a function of ${\rho}$ within
the physically acceptable region.

If $\kappa$ is of order unity, then we can interpret this effect as a
cutoff in energy density
at around the Planck scale. Given that $\rho\propto a^{-4}$ (recall that the conservation equation is the same as in GR, so this relation does not change in this theory), this means that
there is a minimum
value for the scale factor at $a_B\sim 10^{-32} {\rm} (\kappa)^{1/4}a_0$
(where $a_0$ is the
scale factor today)- corresponding to a  {\it minimum length}, $a_B\equiv(\rho_0/\rho_B)^{-4}$, in cosmology.

The nature of the expansion rate at $a_B$ depends on the sign of $\kappa$ as we show in Figure \ref{aoft} where
we plot the scale factor, $a$, for both situations.
 For $\kappa<0$, one
can show that $H^2\propto (a-a_B)$ which means that $a-a_B\propto|t-t_B|^2$ for a fixed
$t_B$.  This means that at $t_B$ the Universe undergoes a regular bounce. The transition through the bounce occurs on a time
scale of $\Delta t\simeq\sqrt{\kappa}$. Again, if $\kappa$ is if order unity, the deviation from
conventional cosmology occurs at the Planck time on a time scale of that order.

A far more interesting, in our view, behavior can be
found if $\kappa>0$ where one can show that $H^2\sim (a-a_B)^2$ which means that $\ln (a/a_B-1)\propto t-t_B$. In this case there is no bounce; if we wind back the clock, the energy density
will reach a point (corresponding to about $\rho_B/2$ as can be seen from Figure \ref{H2rho}) in
which accelerated expansion kicks in. As above, this corresponds to what we would perceive as the Planck time, but now,
the scale factor will take an infinitely long time to reach $a_B$. This leads to an intriguing alternative cosmology at early times in which, depending on how close its
initial density is to $\rho_B$, the Universe loiters for a long time (in terms of Planck units) until
it emerges into a standard cosmological evolution. This evolution can be seen as an alternative incarnation of the Einstein
static universe at early times. As in the conventional construction, the loitering phase is unstable
to expansion.

There are two effects that may be of importance in the early Universe. First of all, it should be clear that we have been looking at the
classical behaviour of this theory of gravity. The onset of a bounce or minimum length may signal pathologies at the quantum
level, such as the presence of ghosts and negative normed states. A more detailed analysis of the gravitational Born-Infeld action
will allow us to check if these pathologies indeed exist. Secondly, we have discarded the effect of $\Lambda$ (i.e. $\kappa\Lambda\ll1$) in our analysis of
the Early Universe. If, however, there is an early time contribution to $\Lambda$ from, for example, an inflaton, the dynamics will
be different; with $\kappa\Lambda\simeq 1$ there will be another inflection point and $H^2(\rho)$ will never reach $0$ for a finite
density. Such scenario merits further analysis.

Eddington's theory of gravity, through its descendant, the gravitational Born-Infeld action, may lead
to an entirely new view of the Universe. As pointed out above, the presence of a maximum density in cosmology may have a bearing not only on the early universe but in the dynamical formation of black holes. Thus, if a mimimum length also arises during gravitational collapse, the Universe may be entirely singularity free \cite{Brandenberger}, solving one of the problems that has troubled relativists since Einstein
first proposed his theory of gravity.


\vspace{-0pt}
\section*{Acknowledgements}
\vspace{-10pt}

We are grateful to T.Clifton, G. Gibbons, C. Skordis, T. Sotiriou and T. Jacobson for useful conversations.  This work was supported by the BIPAC. MB was partially supported by Alma-Conicyt Grant \#  31080001; Fondecyt Grant \#1060648; and the J.S. Guggenheim Memorial Foundation.



\begin{thebibliography}{100}

\bibitem{Eddington} A.S. Eddington, The Mathematical Theory of Relativity,
Cambridge University Press  (1924);  E. Schrodinger, Spacetime Structure,
Cambridge University Press (1950).

\bibitem{kijowski} M. Ferraris and J. Kijowski, Letters in Mathematical
Physics, 5 127-135, (1981);
N. Poplawski, Int.J.Mod.Phys.D18:809-829 (2009); Found.Phys.39:307-330
(2009).

\bibitem{comment} If the Hessian $ \delta^2 I/\delta g_{\mu\nu}\delta g_{\alpha\beta} $ is not invertible then this step is not possible. For example, for a massless free boson in flat space the equation of motion is $R_{\mu\nu}=\partial_\mu\phi\partial_\nu\phi$ which does not depend on the metric. These are isolated cases.

\bibitem{BornInfeld}M.~Born and L.~Infeld, Proc. Roy. Soc. {\bf A144}, 425 (1934).

\bibitem{DeserGibbons} 
  S.~Deser and G.~W.~Gibbons,
  Class.\ Quant.\ Grav.\  {\bf 15}, L35 (1998)
  [arXiv:hep-th/9803049].

\bibitem{Vollick}
  D.~N.~Vollick,
  Phys.\ Rev.\  D {\bf 69} (2004) 064030
  [arXiv:gr-qc/0309101].
  D.~Vollick, Phys. Rev. {D72}, 084026 (2005).
  D.~N.~Vollick,
  arXiv:gr-qc/0601136.


\bibitem{other}
  D.~Comelli and A.~Dolgov,
  JHEP {\bf 0411} (2004) 062
  [arXiv:gr-qc/0404065].
  J.~A.~Nieto,
  Phys.\ Rev.\  D {\bf 70} (2004) 044042
  [arXiv:hep-th/0402071].
  D.~Comelli,
  Phys.\ Rev.\  D {\bf 72} (2005) 064018
  [arXiv:gr-qc/0505088].
  Yu.~M.~Zinoviev,
  arXiv:hep-th/0504210.
  Yu.~M.~Zinoviev,
  arXiv:hep-th/0506217.
  D.~Comelli,
  J.\ Phys.\ Conf.\ Ser.\  {\bf 33} (2006) 303.
  R.~Ferraro and F.~Fiorini,
  Phys.\ Rev.\  D {\bf 78} (2008) 124019
  [arXiv:0812.1981 [gr-qc]].
  R.~Ferraro and F.~Fiorini,
  arXiv:0910.4693 [Unknown].
  F.~Fiorini and R.~Ferraro,
  Int.\ J.\ Mod.\ Phys.\  A {\bf 24} (2009) 1686
  [arXiv:0904.1767 [gr-qc]].

\bibitem{Sotiriou}
  T.~P.~Sotiriou and V.~Faraoni,
  arXiv:0805.1726 [gr-qc].



\bibitem{Demianski} 
  M.~Demianski,
  Found.\ Phys.\  {\bf 16} (1986) 187.

\bibitem{GibbonsRasheed} 
  G.~W.~Gibbons and D.~A.~Rasheed,
  Nucl.\ Phys.\  B {\bf 454}, 185 (1995)
  [arXiv:hep-th/9506035].

\bibitem{Brandenberger} M.~Markov, Pis'ma Zh. Eksp. Theor. Fiz {\bf 36}, 214 (1982);
M.~Markov, Pis'ma Zh. Eksp. Theor. Fiz {\bf 46}, 342 (1987); V.~Mukhanov and R.~Brandenberger, Phys. Rev. Lett. {\bf 68}, 1969 (1992).

\end{thebibliography}
\end{document}